%% file: main.tex
\documentclass[a4paper, amsfonts, amssymb, amsmath,  showkeys,superscriptaddress, nofootinbib, aps,  twocolumn, prl, floatfix]{revtex4-2}
\usepackage[english]{babel}
\usepackage[utf8]{inputenc}
\usepackage[colorinlistoftodos, color=green!40, prependcaption]{todonotes}
\usepackage{dsfont}
\usepackage{amsmath,amssymb}
\usepackage[normalem]{ulem}
\input{preamble}
\usepackage[pdftex, pdftitle={Article}, pdfauthor={Author}]{hyperref} % For hyperlinks in the PDF

\makeatletter
\DeclareFontEncoding{LS1}{}{}
\DeclareFontSubstitution{LS1}{stix}{m}{n}
\DeclareMathAlphabet{\mathscr}{LS1}{stixscr}{m}{n}
\makeatother

\begin{document}
%\title{$p$-orbital origin spin-spiral state and magnetically-induced polarization in spin-orbital entangled alkali superoxide CsO$_2$}
\title{Entangled orbital, spin, and ferroelectric orders in $p$-electron magnet CsO$_2$}
%\title{Magnetically-induced ferroelectricity in CsO2 via Orbital Ordering}

\author{Ryota Ono}
\email[Correspondence email address: ]{ryota.ono.gm@gmail.com} 
\affiliation{National Institute for Materials Science, MANA, 1-1 Namiki, Tsukuba, Ibaraki 305-0044, Japan}

\author{Ravi Kaushik}
\affiliation{Quantum Materials Theory, Italian Institute of Technology, Via Morego 30, 16163 Genova, Italy}

\author{Sergey Artyukhin}
\affiliation{Quantum Materials Theory, Italian Institute of Technology, Via Morego 30, 16163 Genova, Italy}

\author{Martin Jansen}
\affiliation{Max-Planck-Institut f\"{u}r Festk\"{o}rperforschung, D-70569 Stuttgart, Heisenbergstr. 1, Germany}

\author{Igor Solovyev}
\email[Correspondence email address: ]{SOLOVYEV.Igor@nims.go.jp}
\affiliation{National Institute for Materials Science, MANA, 1-1 Namiki, Tsukuba, Ibaraki 305-0044, Japan}

\author{Russell A. Ewings}
\email[Correspondence email address: ]{russell.ewings@stfc.ac.uk}
\affiliation{ISIS Pulsed Neutron and Muon Source, STFC Rutherford Appleton Laboratory, Harwell Campus, Didcot, Oxon OX11 0QX, United Kingdom}

\date{\today} 

\newpage

\begin{abstract}
%Magnetic alkali superoxides are different from conventional transition metal magnets, as they exhibit magnetism arising from partially occupied oxygen molecular $\pi^*$-orbitals. 
Alkali superoxides differ from conventional transition metal magnets, exhibit magnetism from partially occupied oxygen molecular $\pi^*$-orbitals. 
Among them, CsO$_2$ stands out for its potential to exhibit novel quantum collective phenomena, such as an orbital order induced Tomonaga-Luttinger liquid state.
Using ab-initio Hubbard models, superexchange theory, and experimental spin wave measurements, we propose that CsO$_2$ exhibits unconventional magnetoelectric characteristics at low temperature.
%, where magnetism and ferroelectric polarization are entangled through orbital ordering.
Our analysis confirms a canted antiferromagnetic ground state and a spin-flop transition, with ferroelectricity is induced by breaking inversion and time-reversal symmetry in the spin-flop phase.
Consequently, our analysis reveals a strong interplay not only between exchange interactions but also among magnetically-induced polarization and orbital order.
The magnetic structure, stabilized by orbital order, induces magnetically-induced polarization through an antisymmetric mechanism.
%The ferroelectricity arises from the magnetic structure, which is stabilized by the peculiar orbital order.
%This magnetic structure, in turn, induces the antisymmetric mechanism of the magnetically-induced polarization.
%Consequently, our calculations represent a ferroelectric polarization analog of the well-known Goodenough-Kanamori-Anderson rule for exchange interactions.
%Our investigations, covering both theoretical predictions and experimental observations, provide robust evidence for these phenomena originated by the orbital order in CsO$_2$. 
%Our combined theoretical and experimental investigations provide a robust evidence for these phenomena, positioning 
Overall, our results reveal the coexistence of three highly entangled orders in CsO$_2$, namely, orbital, spin and ferroelectricity.
%These combined theoretical and experimental findings establish CsO$_2$ as a novel multiferroic material, highlighting the crucial role of orbital-order on its magnetic and electronic behaviors, and open new avenues for exploring the complex physics of multiferroic molecular orbital magnets. 
%These findings open new avenues for exploring the complex physics of multiferroic molecular orbital magnets.
%originated by the orbital order in CsO$_2$. 
%Our findings not only establish CsO$_2$ as a new multiferroic material, but also emphasize the substantial influence of orbital properties on its magnetic and electronic behaviors, thereby opening novel avenues of exploration into the complexities of multiferroic physics of molecular orbital magnets.
\end{abstract}

%\keywords{first keyword, second keyword, third keyword}

\maketitle
%\ro{Let's write it in the format of Nat. Mat. or Nat. Phys.}

%\re{Hi Ryota, apologies for the slow progress on this. I have been getting increasingly good agreement between LSWT fits and the data, so I am hoping to get a definitive fit soon. One interesting thing that I've found is that if I just include the isotropic exchange and the DM terms, I get quite large values for $dz_{\alpha}$, particularly on the F1 bond. This may be an artefact of missing out the $\Gamma$ terms though. On the latter, I notice that towards the end of the paragraph that starts with the words "From the symmetry constraint,..." you define the terms in the Hamiltonian as $\Gamma_{ij}^{xy}(S_i^{x} S_j^{x}+S_i^{y} S_j^{y}) + \Gamma_{ij}^{z} S_i^{z} S_j^{z} $. Should this not be written as the products $S_i^{x} S_j^{y}+S_i^{y} S_j^{x}$? Or have I misunderstood? As defined currently these are perturbations on the values of the diagonal terms in the exchange tensor set the the values of $J_{\alpha}$.}

%\re{OK, thanks for clarifying that - I was confused by the description of equation 3 where you refer to the last term as being traceless. I had originally (wrongly) interpreted this as meaning the diagonal elements were zero, whereas now I think about it I can see that the sum of the values listed in table I of $2 \Gamma_{||} + \Gamma_{z} = 0$, which also gives a traceless matrix.}

%\ro{Calculated LSTW result looks great. Thank you!}
Understanding the origins and mechanisms underpinning novel magnetic phenomena in transition metal oxides has been at the forefront of condensed matter physics research. 
Particularly, the interplay among spin, orbital, and lattice degrees of freedom, and a possibility to utilize this triad for controlling and designing material properties has been the focus over the past decades. The transition-metal oxides were regarded as the key materials in this context~\cite{PhysRevX.10.031043,KHOMSKII202498,PhysRevB.72.214431,PhysRevLett.102.017205}. However, it appears that there is another class of materials, exhibiting similar properties without traditional $d$ or $f$ elements. These are the alkali superoxides $A$O$_2$ ($A$ = K, Na, Cs, or Rb). The oxygen molecule, O$_2$, is one of the few examples of the molecules, forming the ``high-spin'' triplet ground state, in analogy with atomic states satisfying Hund's rules. Therefore, substances containing O$_2$ molecules are potentially magnetic. 
%\ro{In this class, the magnetism is coming from the partially occupied \emph{molecular} $p$-states of the charged O$_{2}^{-}$ complexes instead from partially occupied atomic $d$ orbitals in transition-metal oxides.} 
%If magnetism of the traditional-metal oxides is related to the partially occupied atomic $3d$ shell, the magnetism of $A$O$_2$ is due to the partially occupied \emph{molecular} $p$-states of the charged complexes O$_{2}^{-}$, forming the lattice. 
While the magnetism in traditional transition-metal oxides is linked to the partially occupied atomic $3d$ shell, in $A$O$_2$ compounds, magnetism stems from the partially occupied molecular $p$-states of the charged O$_{2}^{-}$ complexes that form the lattice.
Moreover, the electronic structure of O$_{2}^{-}$ dimer is such that the active states are the four-fold spin-orbital degenerate molecular $\pi^{*}$ states, which accommodate three electrons. Thus, we have a canonical example of the $S=1/2$ spin system, which is nevertheless prone to orbital effects, depending on how the degeneracy of the $\pi^{*}$ states is lifted~\cite{NaO2,PhysRevLett.108.217206,PhysRevLett.115.057205,Solovyev_2008,PhysRevB.102.085129}.
%\sout{Alkali superoxides offer an unique platform for investigating such phenomena, owing to their intricate interplay of spin, orbital, and lattice interactions}~\cite{NaO2,PhysRevLett.108.217206,PhysRevLett.115.057205,Solovyev_2008,PhysRevB.102.085129}. \sout{Notably, the magnetism in alkali superoxides originates from the $p$-orbital of oxygen, makes them peculiar example. In these materials, two oxygen atoms form a molecule within the solid with a 3/4-filled electron configuration in the antibonding $\pi^*$ orbital, facilitating the $S=1/2$ spin states.}

At high temperatures, the alkali superoxides adopt either a cubic or a tetragonal phase~\cite{Dudarev1973}, so that the molecular $\pi^{*}$ orbitals remain degenerate. However, when lowering the temperature, many of them exhibit a structural transition to a lower symmetry phase, typically an orthorhombic one  \cite{RbO2,Nakano_2023,Russell}, which breaks the degeneracy and results in the $\pi^{*}$ molecular orbital ordering in a specific way. Due to the shape of the $\pi^{*}$ molecular orbitals, the key principles of the spin order, emerging in response to this orbital order, can deviate substantially from those established for the conventional atomic $d$ or $f$ orbitals.  
%\sout{At the room temperature, certain alkali superoxides, such as RbO$_2$ and CsO$_2$, crystallize in the tetragonal structure, characterized by a fourfold rotational symmetry about the crystal $c$-axis~\cite{Dudarev1973}. 
%This symmetry leads to the complete degeneracy of $p_x$ and $p_y$ molecular orbitals representing the $\pi^*$ state. 
%However, at low temperatures, this four-fold symmetry can be broken by lattice distortions, leading to an orbital ordering.
%This molecular orbital-driven magnetism deviates significantly from conventional systems where magnetism primarily arises from the $d$ or $f$ orbitals.} 
Recent investigations have unveiled the potential for exotic states in alkali superoxides, driven by the interplay between orbital order and magnetic interactions~\cite{NaO2,PhysRevLett.108.217206,PhysRevB.109.235115}.

Ferroelectric polarization is another principal degree of freedom. All known alkali superoxides adopt centrosymmetric structures in the paramagnetic state and, therefore, are expected to be paraelectric. 
Nevertheless, a very important question is whether the inversion symmetry can be broken by a peculiar magnetic order,  either spontaneously, as was observed in multiferroic transition-metal compounds~\cite{Kimura2003}, or by magnetic field, which is the essence of the magnetoelectric effect. 
Theoretical analysis suggests that at least one alkali superoxide, NaO$_2$, could become multiferroic~\cite{C3CE41349G}. 
However, such analysis strongly relies on the experimental crystal structure, which controls the  orbital ordering and the behavior of interatomic exchange interactions.
%\sout{Additionally, there is another principal degree of freedom, ferroelectric polarization.
%Since alkali superoxides exhibit a Mott insulating state, it is natural that complex orbital-order induced magnetism can destroy the centrosymmetry of the material and generate finite polarization, as seen in several materials~\cite{Kimura2003}. 
%Indeed, one of the alkali superoxides, NaO$_2$, is suggested to exhibit the potential for finite electric polarization~\cite{C3CE41349G}.}

CsO$_2$, another alkali superoxide, has shown promise for exhibiting $\pi^*$ orbital order-driven low-dimensional magnetism, including the possibility of a Tomonaga-Luttinger liquid state~\cite{PhysRevLett.108.217206,PhysRevLett.115.057205}. The canonical mechanism describing the coupling of spins to the crystal structure of alkali superoxides is believed to be magnetogyration, associated with the rotational degrees of freedom of the O$_2^-$ molecules~\cite{Magnetogyration,KO2_chem,PhysRevLett.108.217206}.
However, recent analysis of the low-temperature $Pnam$ orthorhombic phase of CsO$_2$ suggests a different scenario, with Jahn-Teller distortions involving Cs atoms lifting the orbital degeneracy~\cite{Russell}.
Although there has been an experimental study on the low temperature crystal and magnetic structure of CsO$_2$, the underlying microscopic mechanisms have not yet been elucidated.

%This paper aims to explore the intricate properties of CsO$_2$ by utilizing advanced theoretical models and experimental technique to unravel the interplay between spin and orbital degrees of freedom, elucidating their implications for both magnetic and electric properties.
%Our microscopic investigation of the superexchange mechanisms and their impacts on the magnetic and electric order in CsO$_2$ highlights its potential as an ideal model system for exploring the complex interplay among spin, orbital, and electric degrees of freedom. 
%Additionally, our realistic simulations demonstrate the potential for manipulating these degrees of freedom using external fields, highlighting the significance of alkali superoxides as a versatile platform for investigating emergent phenomena in condensed matter physics.

\begin{figure*}[htbp]
\centering
\includegraphics[keepaspectratio, scale=0.13]{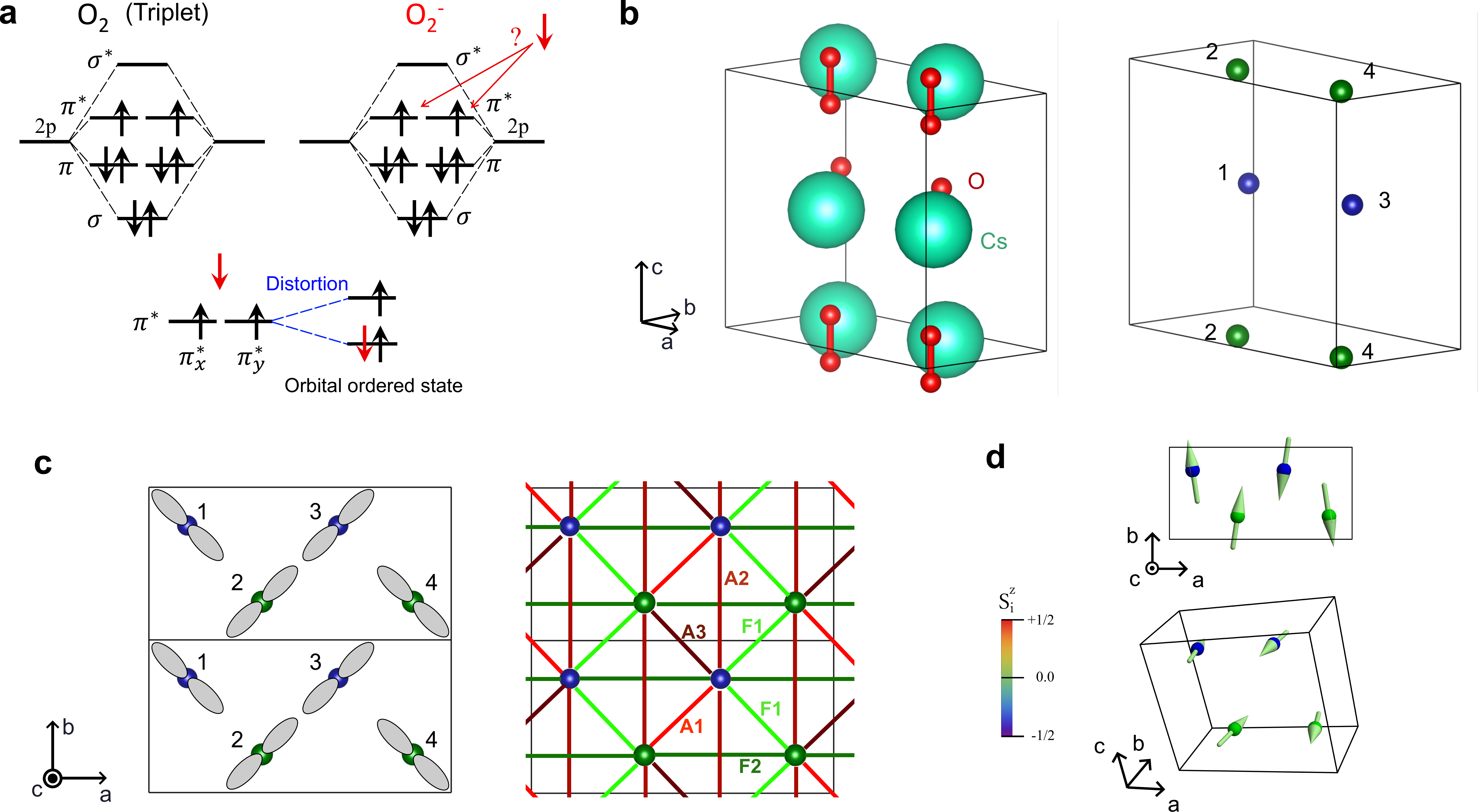}
\caption{\textbf{Orbital and magnetic order in CsO$_2$.} \textbf{a}, A schematic of the electronic configuration in O$_{2}$ and O$_{2}^{-}$ molecular orbitals. Red arrows indicate the electrons, one of which should be removed in O$^{-}_{2}$ in comparison with the electron configuration of the neutral oxygen molecule. Structural distortions destroy this two-fold degeneracy, resulting in a $S=1/2$ orbital singlet state on each O$_{2}^{-}$ molecule. \textbf{b}, Experimentally observed low-temperature orthorhombic crystal structure. Green and blue spheres indicate centers of the O$_{2}^{-}$ molecules, located in different layers. \textbf{c}, Hole orbital-order density $|w_i(\mathbf{r})|$ and geometry of the exchange interactions as obtained from the two-orbital Hubbard model and superexchange theory. \textbf{d}, Canted AFM ground state in CsO$_2$, obtained by the energy minimization of the realistic spin-Hamiltonian with the model parameters in Table \ref{tab:param2}. }\label{fig:fig1}
\end{figure*}

%\textbf{Superexchange interactions} 

%To construct low energy model for CsO$_2$, we employ Wannier downfolding approach for ab initio electronic structure.
%We employ the recently reported orthorombic crystal structure and take spin-orbit coupling (SOC) effect in the electronic structure calculation.
%Then, we construct 2-orbital model for $\pi^*$ orbital, that has hardly dispersive electronic structures located near the Fermi level (Fig.~\ref{fig:bands}). 
%Coulomb interactions are calculated by constrained RPA method for the downfolded model, and estimated as Kanamori parameters $U$=6.8~eV and $J_H$=0.5~eV, thus the strong coupling limit superexchange theory is considered to be valid.
%The resulting realistic 2-orbital Hubbard model is given as 
%To understand physics in CsO$_2$ correctly, it is necessary to perform realistic calculation with 
In alkali superoxides, the molecular orbitals of each O$_2^-$ molecule are, in ascending order of energy: $\sigma$, $\pi$, $\pi^*$, and $\sigma^*$.
These energy levels are strongly split, allowing for a reduction in the Hilbert space.
Consequently, the O$_2^-$ electronic configuration results in a 3/4-filling of the $\pi^*$ orbital as shown in Fig.~\ref{fig:fig1}\textbf{a}.
Therefore, it is valid to restrict a model of the system to only the $\pi^*$ orbital.
Utilizing the recently reported orthorhombic crystal structure ~\cite{Russell}, shown in Fig.~\ref{fig:fig1}\textbf{b}, we incorporate the effects of spin-orbit coupling (SOC) in our electronic structure calculations. 
In our setting, the lattice vectors are taken as $\mathbf{a}=a\hat{x}, \mathbf{b}=b\hat{y}, \mathbf{c}=c\hat{z}$ ($a=8.7271$ \AA, $b=4.3976$ \AA, $c=7.3386$ \AA).
Subsequently, we construct a realistic two-orbital Hubbard model for the $\pi^*$ orbital, characterized by its nearly dispersionless electronic states, located near the Fermi level (Supplementary Fig.~1).
The Hamiltonian has the form:
\begin{align}
\label{eqn:Hamiltonian}
    \mathcal{H} = \mathcal{H}_{\text{1el}}+\mathcal{H}_{\text{K}}+\mathcal{H}_{\text{V}},
\end{align}
where $\mathcal{H}_{\text{1el}}$ is the one-electron part, $\mathcal{H}_{\text{K}}$ are the intra-molecular Coulomb interactions, and $\mathcal{H}_{\text{V}}$ denotes the inter-molecular Coulomb interactions, respectively:
\begin{align}
\label{eqn:eachterms}
    & \mathcal{H}_{\text{1el}} = \sum_{ij} \sum_{ab\sigma\sigma'} t_{ij}^{ab\sigma\sigma'} \hat{c}_{ia\sigma}^\dagger \hat{c}_{jb\sigma'}, \\
    & \mathcal{H}_{\text{K}} = U \sum_{ia} \hat{n}_{ia\uparrow} \hat{n}_{ia\downarrow} + U' \sum_{ia > b} \hat{n}_{ia} \hat{n}_{ib} \notag \\
        & \qquad - J_H \sum_{i\sigma\sigma' a > b} \hat{c}_{ia\sigma}^\dagger \hat{c}_{ib\sigma'}^\dagger \hat{c}_{ia\sigma'} \hat{c}_{ib\sigma} \notag \\
        & \qquad + J_H \sum_{ia > b} \left( \hat{c}_{ia\uparrow}^\dagger \hat{c}_{ia\downarrow}^\dagger \hat{c}_{ib\downarrow} \hat{c}_{ib\uparrow} + \text{h.c.} \right), \\
    & \mathcal{H}_{\text{V}} = \sum_{ijab} V_{ijab} \hat{n}_{ia} \hat{n}_{jb},
\end{align}
where $\hat{c}_{i a \sigma} ^{\dagger}$ ($\hat{c}_{i a \sigma}$) stands for the creation (annihilation) of an electron on the Wannier orbital $a$ of the molecule site $i$ with the spin $\sigma$, and $\hat{n}_{i a \sigma} = \hat{c}_{i a \sigma} ^{\dagger} \hat{c}_{i a \sigma}$ is the number operator.
Here, $U$ is the intra-orbital Coulomb interaction, while the inter-orbital Coulomb interaction $U' = U-2J_{H}$ is reduced due to the Hund's coupling $J_H$.
The Coulomb interactions are evaluated using the constrained random phase approximation (cRPA)~\cite{crpa} applied to the downfolded model, yielding estimated Kanamori parameters of $U=6.8$ eV, $J_H=0.53$ eV and $V_{ijab}=1.3-1.4$ eV for all main bonds shown in Fig.~\ref{fig:fig1}\textbf{c}, respectively.
Thus, the model operates within the strong coupling regime, validating the application of superexchange theory. 

The resulting orbital order (the shape of unoccupied $\pi^*$ orbitals obtained from the diagonalization of $[t_{ii}^{ab\sigma\sigma'}]$ at each molecular site $i$) is  shown in Fig.~\ref{fig:fig1}\textbf{c}.
It features a two-tetragonal-cell-period structure along the crystal $a$-axis and a single-tetragonal-cell-period structure along the crystal $b$-axis. 
We note that the parallel hole-orbital alignment at the middle of the unit cell hints at a strong antiferromagnetic (AFM) exchange interaction.
%This can arise from the fact that the only one of the doubly degenerate $\pi^*$ orbital can be half-filled.
%The orbital-order shows two tetragonal cell period orbital order structure along crystal $a$-axis, while single tetragonal cell period along crystal $b$-axis.
%The observed orbital order exhibits a two-tetragonal-cell-period structure along the crystal $a$-axis and a single-tetragonal-cell-period structure along the crystal $b$-axis. 
%Parallel orbital order at the middle of the unit cell indicates a strong antiferromagnetic exchange interaction.

\begin{comment}

\begin{figure}[htbp]
\centering
\includegraphics[keepaspectratio, scale=0.18]{./BANDS_PDOS.png}
\caption{Electronic band structure from DFT and 2-orbital model for Oxygen $\pi^*$ orbital.}\label{fig:bands}
\end{figure}

\end{comment}

By employing the superexchange theory for the model (\ref{eqn:Hamiltonian}), the low-energy Hamiltonian is then mapped to the general spin-Hamiltonian for $S=1/2$ systems, given as:
\begin{align}
 \label{eqn:H_s}
    \mathcal{H}_s = &\sum_{<ij>} \left( J_{ij} \textbf{S}_i \cdot \textbf{S}_j+ \textbf{D}_{ij} \cdot (\textbf{S}_i \times \textbf{S}_j) + \textbf{S}_i \cdot \stackrel{\leftrightarrow}{\Gamma}_{ij} \textbf{S}_j \right),
\end{align}
%\com{Shall we use more conventional $\textbf{S}_i$ instead of $\textbf{e}_i$ and, therefore, multiply all numerical values for the exchange interactions by $1/S^2=4$?}
%where the first term is isotropic exchange, the second term is Dzhalishonskii-Moriya (DM) interaction and the last term is the symmetric anisotropic exchange interaction for spin pair at site $i$ and $j$, indicated as $\textbf{e}_i$ and $\textbf{e}_j$, respectively.
where the first term is the isotropic exchange interaction, the second term represents the Dzyaloshinskii-Moriya (DM) interaction, and the last term is the traceless symmetric anisotropic exchange interaction between the spins $\mathbf{S}_i$ and $\mathbf{S}_j$ at sites $i$ and $j$, respectively.
Each term is obtained by decomposing the general $3 \times 3$ spin interaction tensor into symmetric and antisymmetric parts.
In the superexchange process, the intermediate 2-electron excited state is obtained by diagonalizing the on-site Hamiltonian, that consists of SOC, crystal field and the intra-molecular Coulomb interactions.

Furthermore, although the P$nam$ crystal structure is centrosymmetric (non-polar), the inversion symmetry can still be broken by the magnetic structure, which in turn enables the possibility of finite electric polarization.
Utilizing the two-orbital model, we derive the general formula for the magnetically-induced ferroelectric polarization in $S=1/2$ systems~\cite{PRL_IRS,PRB_RIS} given by 
\begin{align}
 \label{eqn:P_s}
    \mathbf{P}_s = &\sum_{<ij>} \left( \textbf{P}_{ij} \textbf{S}_i \cdot \textbf{S}_j +  \stackrel{\leftrightarrow}{\mathbf{\mathcal{P}}}_{ij}\cdot (\textbf{S}_i \times \textbf{S}_j) + \textbf{S}_i \cdot \stackrel{\leftrightarrow}{\mathbf{\Pi}}_{ij} \textbf{S}_j \right).
\end{align}
As in the spin Hamiltonian~(\ref{eqn:H_s}), terms here are the isotropic, antisymmetric and traceless symmetric anisortropic, respectively.
Due to the vector nature of the polarization, these terms are one rank higher compared to the corresponding terms in the spin Hamiltonian.
For instance, the antisymmetric exchange striction coefficients form a $3 \times 3$ tensor, while the corresponding exchange interaction energy coefficients form a DM vector.

\subsection{Results}
\textbf{Spin Hamiltonian and magnetic ground state.}
The resulting spin interactions (AFM in red and FM in green) are illustrated in Fig.~\ref{fig:fig1}\textbf{c} and the exchange parameters are summarized in %Table~\ref{tab:param} and
Table~\ref{tab:param2}, respectively. 
The strongest exchange interaction, $J_{A1}$, is AFM,
and is periodically repeated along the $c$-axis in a zigzag manner.
%The strongest exchange interaction, $J_{A1}$, is AFM interaction structured along the crystal \textbf{c}-axis in a zig-zag bond manner. 
%Such relation of the exchange parameters being AFM ($J_{ij} > 0$) or FM ($J_{ij} < 0$) are tightly bonded to the hole orbital order. 
The alternation of antiferromagnetic ($J_{ij} > 0$) and ferromagnetic ($J_{ij} < 0$), exchange interactions is tied to the hole orbital order: the parallel hole orbital (ferro-orbital, FO) bonds enhance antiferromagnetic exchange interaction, while the anti-parallel hole orbital (antiferro-orbital, AFO) bonds enhance ferromagnetic exchange interaction (see Fig. \ref{fig:fig1}\textbf{c}).
This represents the Goodenough-Kanamori-Anderson (GKA) rules for the exchange interactions~\cite{GK}.
The bonds shown in Fig. \ref{fig:fig1}\textbf{c} can be categorized into two types by their symmetry.
The first type is the centrosymmetric bonds ($A1$ and $A3$), where only the symmetric exchange components, the first and last terms in (\ref{eqn:H_s}), are non-zero.
In the simplest single-orbital case, the exchange interaction would be further restricted to be isotropic~\cite{PRL_IRS}. 
%In these bonds, the isotropic and symmetric anisotropy remain finite.
In these bonds, the symmetric anisotropic interaction is characterized by a parameter $\Gamma_{ij}^{||}$, and the symmetric anisotropic tensor is restricted to $\Gamma_{ij}^{||}(S_i^{x} S_j^{x}+S_i^{y} S_j^{y}) -2 \Gamma_{ij}^{||} S_i^{z} S_j^{z}$.
The second type is the bonds that are connected by a two-fold rotation around the crystal $c$-axis. 
For these bonds, the DM vector is restricted to the $c$-direction.
%Due to the low symmetry structure of this material, several anisotropic exchanges remain finite.
Moreover, the bond $1-3$ (see Fig.~\ref{fig:fig1}\textbf{b}) is transformed into the bond $4-2$ by spatial inversion. 
Therefore, $\mathbf{D}_{1,3}=-\mathbf{D}_{2,4} = (0,0,dz_{F2})$, and both $\mathbf{D}_{1,3}$ and $\mathbf{D}_{2,4}$ are defined by a common parameter $dz_{F2}$.
%In addition, the inversion symmetry between the $1-3(0,0,0)$ and $4-2(0,0,0)$ bonds enforces the DM vector to alternate in sign. 
%Then, since the bond $1-3(0,0,0)$ and $4-2(0,0,0)$ are connected by inversion symmetry, the DM vector alternates the sign.
%For instance, $\mathbf{D}_{1,3(0,0,0)} = (0,0,dz_{F2})$ and $\mathbf{D}_{2,4(0,0,0)} = (0,0,-dz_{F2})$ are both represented by a common parameter $dz_{F2}$.
The symmetric anisotropic tensor for these bonds has the same form as for the centrosymmetric bonds.
%On the other hand, the bond $A1$ and $A3$ are on the inversion center.
%This restricts the spin-Hamiltonian to be only isotropic in these bonds~\cite{PRL_IRS}.

Minimization of the full spin Hamiltonian~(\ref{eqn:H_s}) results in a single cell-periodic canted AFM structure, depicted in Fig.~\ref{fig:fig1}\textbf{d}.
%Minimization of the full spin-Hamiltonian Eq.~(\ref{eqn:H_s}) reveals the spin-spiral ground state shown in Fig.~\ref{fig:GS} with three-cell period spiral characterized by $q \approx (0, 2/3, 0)$.
%Owing to the high-frustration of this material, the magnetic structure is stabilized with a spiral state.
%The spin-spiral plane is fixed within crystal $\mathbf{ab}$-plane, primarily can be considered due to several DM vectors pointing along crystal $\mathbf{c}$-vector.
%This is slightly disagreeing with the experimental ground state reported in Ref.~\cite{Russell}, which is a canted AFM with spins in $\mathbf{ab}$-plane.
%However, the second dataset (Table~\ref{tab:param2}) yields a ground state very similar to the experimental ground state (Fig.~\ref{fig:GS2}).
The spins lie within the crystallographic $\mathbf{ab}$-plane, which is principally due to the DM vectors oriented along the $c$-axis.
%primarily can be considered due to several DM vectors along the crystal $\mathbf{c}$-vector.
The obtained spin structure agrees well with the experimental ground state~\cite{Russell}.
The calculated transition temperature is approximately 8\,K (Supplementary Fig.~2), %(Fig.~\ref{fig:Cv}), 
also in a good agreement with the experimental value 10\,K~\cite{Russell}.

\begin{table}
    %\captionof{table}
    \caption{Values of exchange parameters for each bond [meV] %calculated with 
    %$U=3.55$~eV and $J_H=0.62$~eV~\cite{Solovyev_2008}
}\label{tab:param2}
\begin{ruledtabular}
\begin{tabular}{cccc}
 $\alpha: (i, j, \mathbf{R}_j- \mathbf{R}_i)$ & $J_\alpha$ & $dz_\alpha$ & $\Gamma_\alpha^{||} $  \\
\hline
 A1: $\left(2,3  \right)$ & 3.067 & & 0.499  \\
 A2: $\left(1,1, \mathbf{b} \right)$ & 0.653 & 0.081 & -0.002 
\\  
 A3: $\left(3,2,\mathbf{b} \right)$ &  0.265 & 
 &0.487  \\
 F1: $\left(3,4  \right)$ &  -1.548 &  -0.111 & -0.235  \\  
 F2: $\left(1,3  \right)$ &  -0.209 &  -0.344 & -0.255  \\

\end{tabular}
\end{ruledtabular}
\end{table}

%\re{I will add here a separate section in which the neutron data will go. We can worry about the precise ordering of everything later.}
\begin{figure*}[htbp]
\centering
\includegraphics[keepaspectratio, scale=0.45]{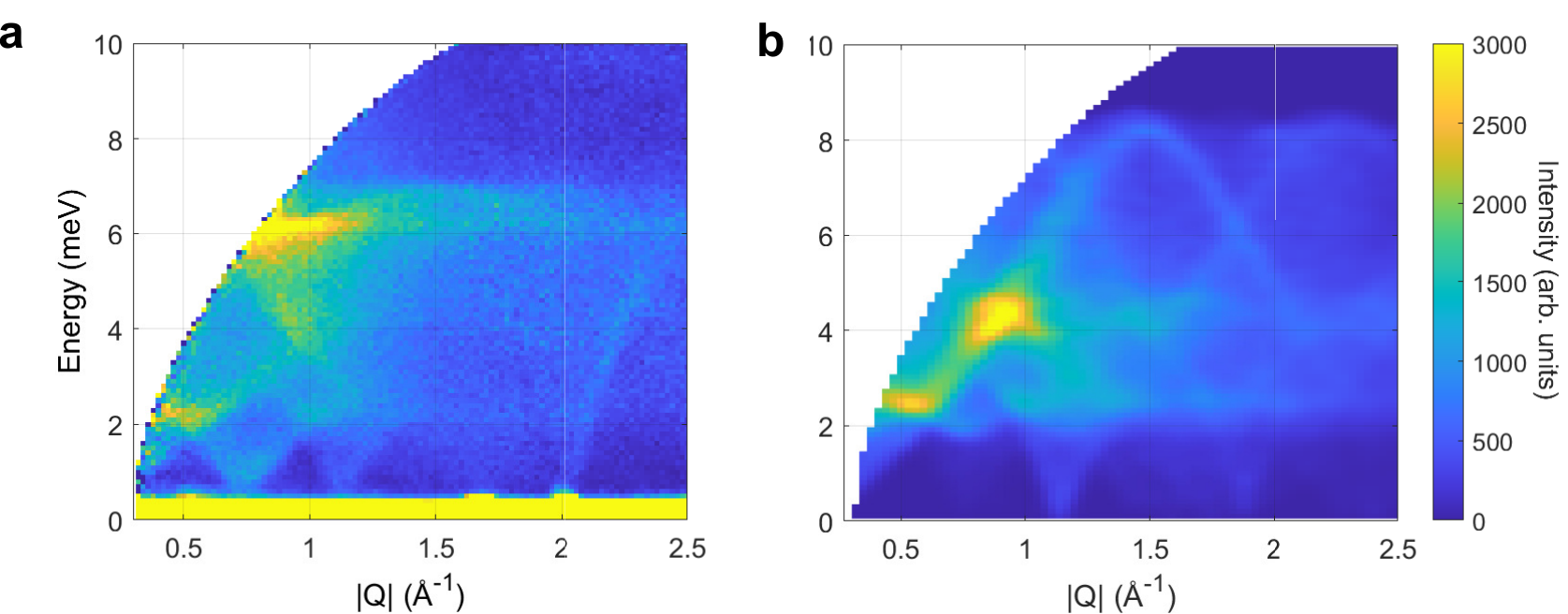}
\caption{\textbf{Experimental and simulated inelastic neutron scattering intensity in CsO$_2$.} \textbf{a}, Inelastic neutron scattering data collected at a temperature of 1.6\,K, in the magnetically ordered phase. The neutron cross-section is indicated by the color map, and is shown as a function of |Q| (modulus of scattering wavevector) and excitation energy. \textbf{b}, Simulated inelastic neutron scattering intensity for the same range in momentum and energy as shown in \textbf{a}, using the model parameters from Table \ref{tab:param2}.}\label{fig:INS}
\end{figure*}

%\subsection{Inelastic neutron scattering}
\textbf{Inelastic neutron scattering results.} 
To understand the microscopic behavior of the magnetism at low temperature in CsO$_2$ and verify our theoretical model, we performed inelastic neutron scattering (INS) measurements of the magnetic excitation spectrum.
Fig.~\ref{fig:INS}\textbf{a} shows the INS data, i.e. $S(\mathbf{Q},\omega)$, collected well into the ground state canted AFM phase, at a temperature of 1.6\,K, as a function of wavevector and energy. The intense and highly structured signal below $\sim 7$\,meV at lower |Q| arises from spin-wave-like excitations, since they appear to follow the expected magnetic form factor. The intense band of scattering below 0.5\,meV corresponds to incoherent elastic scattering, broadened by the instrumental resolution. The sharp dispersive mode centered on the elastic line at $|Q| = 2$\,\AA$^{-1}$ corresponds to an acoustic phonon originating at the strong nuclear Bragg peak (1,1,0) and can be disregarded.

%\re{TODO: Make and include figures showing fits with linear spin wave theory.

%Additional figure (?) of the same data, but with different color saturation, to show the continuum scattering indicative of a spinon continuum.}

\begin{comment}

\begin{figure}[htbp]
\centering
\includegraphics[keepaspectratio, scale=0.55]{./LET_data_colormap.eps}
\caption{Inelastic neutron scattering data collected at a temperature of 1.6\,K, in the magnetically ordered phase. The neutron cross-section is indicated by the color map, and is shown as a function of |Q| (modulus of scattering wavevector) and excitation energy. }\label{fig:ins_data}
\end{figure}
    
\end{comment}

The magnetic excitations visible in the INS data may be analysed using linear spin wave theory (LSWT), using the exchange interactions described in Table~\ref{tab:param2}. Doing so, the colormap shown in Fig.~\ref{fig:INS}\textbf{b} is produced. 
Although this does not reproduce the measured signal shown in Fig. \ref{fig:INS}\textbf{a} especially well, it does capture some of the important features. In particular, it includes a branch of excitations that appear to disperse upwards from $|Q| = 0$. It also reproduces quite well the series of hot spots in intensity along the $|Q|$--axis with an energy of $\sim 2.5$\,meV. Furthermore, the overall bandwidth of the excitations is $\sim 8$\,meV, which is close to the measured one of $\sim 7$\,meV.
Together, this therefore indicates that this set of parameters could serve as a good starting point for fitting the measured neutron spectra, with the expectation that the signs and relative magnitudes of the various terms may not need to change much to yield good agreement. 
This consistency between LSWT results based on our model, and the INS data, comprises clear experimental verification of the calculated orbital order and exchange interactions in CsO$_2$. A more detailed, but ultimately unsuccessful, attempt to fit the data using this spin wave model is provided in the Supplementary Information.
%This is the direct experimental evidence of the calculated orbital order and exchange interactions in CsO$_2$.

\begin{comment}

\begin{figure}[htbp]
\centering
\includegraphics[keepaspectratio, scale=0.55]{./LET_Ryota_pars_sim_colormap.eps}
\caption{Simulated inelastic neutron scattering data for the same range in momentum and energy as shown in fig. \ref{fig:ins_data}, using the model parameters from Table \ref{tab:param2}.}\label{fig:ins_sim_bare}
\end{figure}
\end{comment}

\begin{figure*}[htbp]
\centering
\includegraphics[keepaspectratio, scale=0.30]{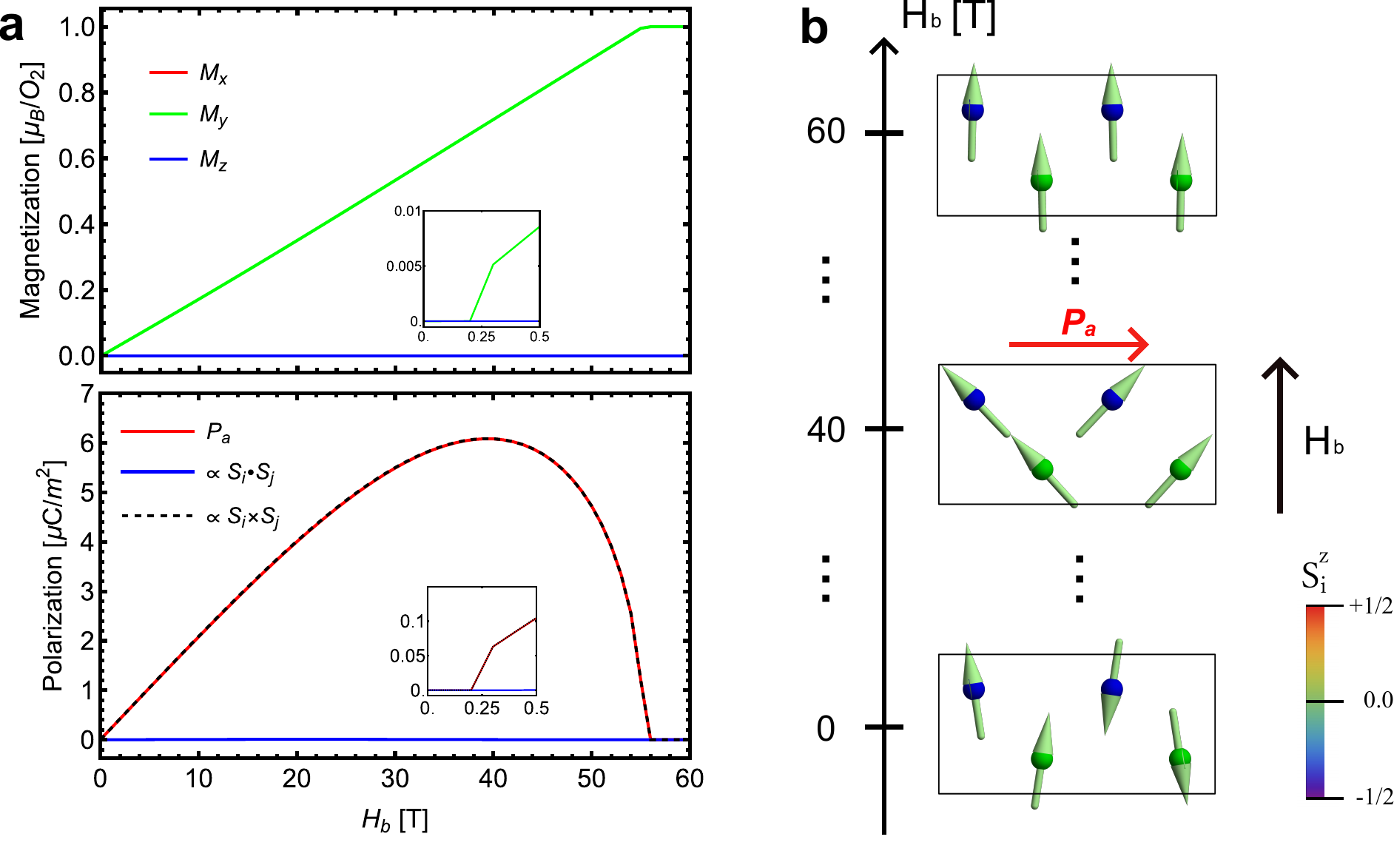}
\caption{\textbf{Magnetic ground state, and magnetic and ferroelectric behavior of CsO$_2$ under an external magnetic field.} 
%\textbf{a}, Real orbital-order (AFM and FO) and their relation to the exchange parameters, in-plane position matrix elements, and the magnetically-induced polarization parameter $p_{ij}$. Two bonds (AFO and FO) from Fig. \ref{fig:fig1}\textbf{c} are chosen to be shown here.
%Green and blue spheres indicate center of the O$_{2}^{-}$ molecules located in different layers. 
\textbf{a}, Magnetization and polarization ($P_a$) changes under an external magnetic field along the crystalline $b$-axis, simulated using the model parameters in Table \ref{tab:param2} and Table \ref{tab:Pparam}. Polarization along other directions, not shown here, remains zero during this process. In the polarization plot, contributions from the isotropic and the antisymmetric mechanisms are shown by solid blue line and dashed black line, respectively. The insets show zoomed plots around the transition field. \textbf{b}, Spin structure evolution under an external magnetic field. The magnetically-induced polarization direction is indicated by a red arrow. In a finite external magnetic field, the polarization keeps the same direction while modulating its magnitude.}\label{fig:mag}
\end{figure*}

%\textbf{Magnetic ground state and magnetically-induced polarization.} 
%hese results suggest the reliability of our theoretical model.
%These suggest that the correct Coulomb interactions to be used here are those reported in~\cite{Solovyev_2008}.
%Therefore, for the later analyses, we employ the second dataset (Table~\ref{tab:param2}).
%The $Pnam$ crystal structure is centrosymmetric (non-polar), however, the resulting spin-spiral structure destroys this inversion symmetry.

\textbf{Magnetically-induced polarization.}
Although the canted AFM ground state preserves the I$\times$T (I=inversion, T=time reversal) symmetry, experiments~\cite{CsO2_Miyajima,Russell} have suggested the possibility of a spin-flop transition induced by an external magnetic field along the $\mathbf{b}$-axis.
During this process, the I$\times$T symmetry can be broken, potentially inducing a finite electric polarization.

The tensors characterizing magnetically-induced polarization  (\ref{eqn:P_s}) have several non-zero elements in all bonds, with their form constrained by the bond symmetry.
A notable example is the centrosymmetric bonds $A1$ and $A3$, where only the antisymmetric exchange striction (the second term in (\ref{eqn:P_s})) contributes. 
Microscopically, this occurs because when the bond is at the inversion center, the hopping integral remains unchanged under the inversion operation, while the position operator changes sign~\cite{PRL_IRS}.
Other noncentrosymmetric bonds can have finite coefficients in all terms in (\ref{eqn:P_s}).
However, as shown in the subsequent results for magnetically-induced polarization, the antisymmetric mechanism is found to be the main contributor, while other mechanisms do not contribute to polarization in our scenario. 
Therefore, in discussing the magnetically-induced polarization, we focus solely on the effects arising from the antisymmetric mechanism. 
In addition, the calculated parameters suggest that finite elements in the antisymmetric exchange striction coefficients for the main bonds given in Fig.~\ref{fig:fig1}\textbf{c} are only $\mathbf{\mathcal{P}}_{\alpha}^{xz}$ and $\mathbf{\mathcal{P}}_{\alpha}^{yz}$.
Therefore, the model for the magnetically-induced polarization (\ref{eqn:P_s}) is reduced to 
\begin{align}
 \label{eqn:P_s_short}
    \mathbf{P}_s \approx &\sum_{<ij>} \left(   \mathbf{\mathcal{P}}_{ij}^{xz}, \mathbf{\mathcal{P}}_{ij}^{yz}, 0 \right) \left( S_i^x S_j^y - S_i^y S_j^x \right).
\end{align}
This means that the magnetically-induced polarization will be in the $\mathbf{ab}$-plane regardless of the magnetic structure.
Furthermore, the parameters are related to the bond directions because of the nature of the position operator.
In fact, all the coefficients in (\ref{eqn:P_s_short}) can be represented as  
\begin{align}
 \label{eqn:P_ij_short}
    \left( \mathbf{\mathcal{P}}_{ij}^{xz}, \mathbf{\mathcal{P}}_{ij}^{yz}, 0 \right)= p_{ij} (\hat{n}^z \times \hat{\epsilon}_{ij}), 
\end{align}
where $\hat{n}^z=(0,0,1)$ and $\hat{\epsilon}_{ij}$ is a unit vector pointing along the bond direction.
The parameters of $p_{ij}$ for each bond are presented in Table~\ref{tab:Pparam}.
It is important to note that in our general theory, the polarization is microscopically derived from the perturbation of the Wannier function with respect to the hopping, similar to the exchange parameters. 
Consequently, the parameter $p_{ij}$ shows a strong correlation with orbital ordering.
%\ro{It is worth noting that, in our general theory, since the polarization is microscopically derived from the perturbation of the Wannier function with respect to hopping, like the exchange parameters, the parameter $p_{ij}$ exhibits a strong correlation with the orbital order.}
Specifically, the FO order gives $p_{ij}>0$, while AFO order gives $p_{ij}<0$.
As a consequence, this behavior is reminiscent of the GKA rule for exchange interactions.
Unlike the exchange interactions, the strength of the magnetically-induced polarization tensors does not strongly depend on the level of orbital overlap.
This is due to the microscopic polarization tensor being first-order in the hopping integrals, while the exchange interaction is second-order.

\begin{table}
\caption{Values of $p_{ij}$ for each bond [$\mu C /m^2$]. %\ro{maybe don't need to use Table} %calculated with $U=3.55$~eV and $J_H=0.62$~eV~\cite{Solovyev_2008}.
}
\label{tab:Pparam}
\begin{ruledtabular}
\begin{tabular}{ccccc}
A1 & A2 & A3 & F1 & F2  \\
\hline
20.5 & 4.8 & 6.5 & -13.5 & -4.4
\end{tabular}
\end{ruledtabular}
\end{table} 

The evolution of magnetization and magnetically-induced polarization, along with the corresponding magnetic structure under the influence of a magnetic field along the crystallographic $b$-axis ($H_b$) as obtained from the energy minimization, is depicted in Fig.~\ref{fig:mag}\textbf{a} and Fig.~\ref{fig:mag}\textbf{b}, respectively.
At zero magnetic field, the canted AFM magnetic structure preserves the I$\times$T symmetry of the material, resulting in no net electric polarization.
As the strength of the magnetic field approaches $H_b \approx 0.3$~T, the magnetic structure undergoes a transition into a spin-flop phase. 
The critical magnetic field is in a qualitative agreement with the experimental value ($2-4$\,T).
In this phase, the spins maintain an AFM vector perpendicular to the external field, while canting towards the direction of the magnetic field.
This state breaks the I$\times$T symmetry of the ground state, leading to the emergence of a finite electric polarization.
This is an example of the canonical magnetoelectric effect, first observed in Cr$_2$O$_3$~\cite{DzyaloshinskiiME}.
The resulting electric polarization is observed to be along the crystallographic $a$-axis, originating solely from the antisymmetric mechanism (see Fig.~\ref{fig:mag}\textbf{a}). 
The disappearance of the isotropic contribution $\propto \mathbf{S}_i \cdot \mathbf{S}_j$ can be explained by the symmetry property of $\mathbf{P}_{ij} = (P_{ij}^{x},P_{ij}^{y},P_{ij}^{z})$ (Supplementary Note 3).
In addition, the disappearance of $P_b$ also arises from the symmetry property of the antisymmetric polarization tensor. 
For instance, the bond $A1$ exhibits a symmetry property $\mathbf{\mathcal{P}}_{2,3}^{xz} = - \mathbf{\mathcal{P}}_{4,1}^{xz}$, due to the bond symmetry $\epsilon_{2,3}^{y} = -\epsilon_{4,1}^{y}$, while $\mathbf{\mathcal{P}}_{2,3}^{yz} = \mathbf{\mathcal{P}}_{4,1}^{yz}$ because $\epsilon_{2,3}^{x} = \epsilon_{4,1}^{x}$.
In the spin-flop phase, the commensurate magnetic structure obeys $\mathbf{S}_4 \times \mathbf{S}_1 = -\mathbf{S}_2 \times \mathbf{S}_3$; thus, only the $x$-component of the polarization, $P_a$, remains finite. 
As the magnetic field strength increases, the spins cant, driving a linear increase in net magnetization along the direction of the field.
Notably, the polarization exhibits a maximum at around 40\,T due to the antisymmetric mechanism of the polarization.
At this point, the angles between spins on $A1$ bond approach approximately 90$^\circ$.
Subsequently, beyond 40\,T, the spins gradually align ferromagnetically, resulting in rapid decrease of the polarization.
This decrease occurs because the FM magnetic structure does not break the inversion symmetry.
The calculated magnetically-induced polarization aligns with the prediction from Katsura-Nagaosa-Balatsky (KNB) formula, given by $\mathbf{P}_s \propto \sum_{ij} \hat{\epsilon}_{ij} \times (\mathbf{S}_i \times \mathbf{S}_j) $~\cite{KNB}. 
The maximal value observed here, $P_a^{\mathrm{max}} \approx 6$ $\mu C/m^2$,
is comparable to the polarization value observed in other $S=1/2$ multiferroics~\cite{CuCl2,CuBr2,CuO}.
%This phenomenon cannot be explained by the widely used Katsura-Nagaosa-Balatsky (KNB) formula~\cite{KNB}. 
%According to the KNB theory, the induced polarization would be predicted to be rotated 45$^\circ$ from the crystal $a$-axis direction.

\subsection{Discussion and outlook}
In this study, we have proposed that the molecular $p$-orbital magnet CsO$_2$ exhibits orbital order, which in turn
%can exhibit the orbital-order phenomena,
effectively controls magnetism and magnetoelectric properties.
Specifically, the microscopic theory has identified the magnetic ground state as a canted AFM state, consistent with experimental results.
The calculated spin Hamiltonian parameters allow to qualitatively reproduce our experimental spin-wave dispersion, demonstrating the accuracy of the theory.
Additionally, the microscopic theory has revealed the GKA rule for the magnetically-induced polarization tensor along with the standard GKA rule for exchange interactions in CsO$_2$.
The ferroelectric polarization is induced along the $a$-axis as a result of broken I$\times$T symmetry through the antisymmetric mechanism of the magnetically-induced polarization in the spin-flop phase, aligning to the prediction from the KNB theory.
Notably, our general theory predicts a different response from the KNB theory under a magnetic field applied along the crystallographic $c$-axis: while the KNB theory anticipates finite polarization due to spin canting along the $c$-axis, our theory yields zero polarization in each bond.
This contrast arises from differences in the antisymmetric exchange striction coefficients, which stem from the symmetry of the $\pi^*$ orbitals.
Nevertheless, the net polarization remains zero in both theories.
In summary, our findings establish alkali superoxides as a playground for a complex interplay between orbital ordering and the peculiar $p$-orbital magnetism.
%We expect similar polarization analogue of the GKA rule exihibits not only in other alkali superoxides, but also in transition metal oxides.
%Finally, we propose that a similar reminiscence of the GKA rule in magnetically-induced polarization might exist not only in other alkali superoxides but also in transition metal oxides with orbital order, such as KCuF$_3$.
\begin{comment}
\begin{figure}[htbp]
\centering
\includegraphics[keepaspectratio, scale=0.52]{./Fig5.pdf}
\caption{Change in magnetization and the magnetically-induced polarization by an external magnetic field along crystal $\mathbf{b}$-axis. 
%\ro{I found that this polarization curve is very similar to $<\mathbf{S}_1 \times \mathbf{S}_2>$ of BEC triplon state in a quantum dimer system TlCuCl$_3$ (Figure 1d of \cite{Kimura2016}).}
}\label{fig:Hb_M_P}
\end{figure}

\begin{figure}
\centering
\includegraphics[keepaspectratio, scale=0.08]{./Fig4.pdf}
\caption{Magnetic structure evolution by an external magnetic field along crystal $\mathbf{b}$-axis. The induced ferroelectricity is indicated by red arrows.}\label{fig:Hb_spin}
\end{figure}
\end{comment}

%\begin{figure}[htbp]
%\centering
%\includegraphics[keepaspectratio, scale=0.55]{./Fig5.pdf}
%\caption{Polarization change by $H_b$ as obtained from Monte-Carlo simualtions.}\label{fig:P_Hb}
%\end{figure}

\subsection{Methods}

\begin{comment}
We derive superexchange energy for 2-orbital model. Those are given by the following formulas:
\begin{align}
E^{(2)}=
&-\frac{1}{U-3 J_H}   |\bra{1 -,i} \hat{t}_{ij}\ket{2-,j}|^2 \nonumber \\
&-\frac{1}{2(U + J_H)}  |\bra{1 -,i} \hat{t}_{ij}\ket{1+,j}|^2 \nonumber \\
&-\frac{1}{2(U - J_H)}  |\bra{1 -,i} \hat{t}_{ij}\ket{1+,j}|^2 \nonumber \\
&-\frac{1}{2(U - J_H)}  |\bra{1 -,i} \hat{t}_{ij}\ket{2+,j}|^2 \nonumber  \\
&-\frac{1}{2(U - 3J_H)}  |\bra{1 -,i} \hat{t}_{ij}\ket{2+,j}|^2 \nonumber \\
&+ (i \leftrightarrow j)
\end{align}
\end{comment}
\textbf{Superexchange theory.}
Since alkali superoxide CsO$_2$ is a Mott insulator, we employ the atomic limit superexchange theory to calculate realistic exchange parameters within the two-orbital Hubbard model.
The second-order perturbation superexchange energy can be decomposed into FM and AFM parts.
These are given as follows:
\begin{comment}
\begin{align}
E^{(2)}_\mathrm{FM}=
&-\frac{1}{U-3 J_H}   \left( |\bra{1 -,i} \hat{t}_{ij}\ket{2-,j}|^2  + (i \leftrightarrow j) \right)
\end{align}
\end{comment}
\begin{align}
E^{(2)}_\mathrm{FM}=
&-\frac{ |\bra{1 -,i} \hat{t}_{ij}\ket{2-,j}|^2  + (i \leftrightarrow j)}{U-V_{ij12}-3 J_H+\Delta}  
\end{align}
and
\begin{comment}
\begin{align}
 \label{eqn:E_AFM}
& E^{(2)}_\mathrm{AFM} =  \nonumber \\
& -\frac{U-2J_H}{(U-J_H)(U-3J_H)}   \left( |\bra{1 -,i} \hat{t}_{ij}\ket{2+,j}|^2  + (i \leftrightarrow j) \right) \nonumber \\
& -\frac{U}{U^2 - J_H^2}  \left( |\bra{1 -,i} \hat{t}_{ij}\ket{1+,j}|^2  + (i \leftrightarrow j) \right),
\end{align}
\end{comment}

\begin{align}
 \label{eqn:E_AFM}
 E^{(2)}_\mathrm{AFM} =  & -\frac{|\bra{1 -,i} \hat{t}_{ij}\ket{2+,j}|^2  + (i \leftrightarrow j) }{2(U-V_{ij12}-3J_H+\Delta)}    \nonumber \\
& -\frac{ |\bra{1 -,i} \hat{t}_{ij}\ket{1+,j}|^2  + (i \leftrightarrow j) }{(1+\alpha_{+}^2)(U-V_{ij11}-J_{H}+\Delta) }   \nonumber \\
& -\frac{|\bra{1 -,i} \hat{t}_{ij}\ket{1+,j}|^2  + (i \leftrightarrow j) }{(1+\alpha_{-}^2)(U-V_{ij11}+\Delta+\delta) } .
\end{align}
Here, indices 1 and 2 denote the orbitals, while the symbols $+$ and $-$ indicate the electron (hole) spin on the half-filled orbital.
$\Delta$ represents the energy splitting of the CF+SOC Hamiltonian, $\delta=\sqrt{\Delta^2 + J_H^2}$ and $\alpha_{\pm} =(\Delta\pm\delta)/J_H$. 
%Additionally, $\alpha_{\pm}$ are defined as $(\Delta\pm\delta)/J_H$.
%where index 1 and 2 denote orbital index, $+(-)$ indicates the electron (hole) spin of the half-filled orbital, $\Delta$ is the energy of CF+SOC, $\delta= \sqrt{\Delta^2+J_H^2}$, $\alpha_{\pm} =\frac{\Delta\pm\delta}{J_H}$, respectively.
The anisotropic exchange parameters are obtained by rotating pseudo spins on sites $i$ and $j$, and taking energy difference~\cite{SE,SE2}.

The magnetically-induced polarization can be derived similarly.
The general formula for the polarization in the solid is given by 
\begin{align}
 \label{eqn:P_gen}
   \mathbf{P} = -\frac{e}{V} \sum_i^{\mathrm{occ}} \bra{w_i|\hat{r}}\ket{w_i},
\end{align}
where $\bra{w_i|\hat{r}}\ket{w_i}$ represents the diagonal elements of the position operator in the Wannier basis, and $V$ is the unit cell volume~\cite{Vdb}. 
Then, the correction to the Wannier function is given by the first-order perturbation.
\begin{comment}
\begin{align}
 \label{eqn:exp}
\ket{w_i} \approx \ket{\alpha +, i} -\sum_j \sum_{M}  \frac{\hat{\mathscr{P}}_{jM}}{E_{jM}}   \ket{\beta -, j} \bra{\beta - , j} \hat{t}_{ji}\ket{\alpha +, i}.
\end{align}
\end{comment}
For instance, the FM channel is represented as
\begin{align}
 \label{eqn:perturbation}
\ket{1-,i}' \approx \ket{1-, i} -\sum_j \frac{ \ket{2+, j} \bra{2+, j} \hat{t}_{ji}\ket{1-, i} }{U-V_{ij12}-3 J_H+\Delta}  .
\end{align}
Substituting this formula into Eq.~(\ref{eqn:P_gen}) yields a pair interaction term similar to Eq.~(\ref{eqn:E_AFM}), where one of the hopping integrals in the square is replaced with the position operator.

%There are several finite element of the intersite Coulomb interaction $V_{i\alpha, j\beta}$, where $\alpha$ and $\beta$ are orbital of the superexchange channel.
%This renormalize the on-site Coulomb interaction $U$ as $U \rightarrow U-V_{i\alpha, j\beta}$.

%Obtained parameters by this approach are following.
\textbf{Electronic structure calculations.}
Density functional theory calculations have been performed for the experimental P$nam$ crystal structure~\cite{Russell}. 
These calculations employed norm-conserving pseudopotentials within the Quantum ESPRESSO package~\cite{QE1}. 
The plane wave cutoff was set to 140~Ry, the Brillouin zone was sampled by a 14 × 14 × 14 Monkhorst–Pack $k$-point mesh.

\textbf{Construction of the Wannier functions.}
The molecular $\pi^*$ orbitals are constructed using Maximally-localized Wannier function method~\cite{Marzari} as implemented in Wannier90~\cite{W90}. 
The calculated position operator matrix elements may have numerical errors due to the numerical finite-difference approximation~\cite{posmat}.
Therefore, the resulting magnetically-induced polarization tensors are symmetrized using symmetry operations of the $Pnam$ space group on the third rank tensor $\stackrel{\leftrightarrow}{\mathbf{P}}_{ij}$.

\textbf{Inelastic neutron scattering.}
Time-of-flight inelastic neutron scattering data were collected on a powder sample of CsO$_{2}$, the same as that used in ref.~\cite{Russell}. The neutron spectra were recorded on the LET spectrometer at the ISIS Pulsed Neutron and Muon Source~\cite{LET}. Data were collected with incident neutron energies of 11.1\,meV, 4.81\,meV and 2.68\,meV. The monochromating chopper system was set up to deliver energy resolutions at the elastic line of 0.36\,meV, 0.12\,meV and 0.055\,meV at full-width-half-maximum (FWHM) respectively. The sample was measured at a temperature of 1.6\,K. Additional data were collected on the MERLIN spectrometer at ISIS \cite{Merlin}, with incident neutron energies of 60\,meV, 30\,meV and 18\,meV. The monochromating chopper system in that case was set up to deliver energy resolutions at the elastic line of 2.5\,meV, 1.0\,meV and 0.55\,meV FWHM respectively. Data were collected with the sample at temperatures of 4\,K, 20\,K and 40\,K.

The raw neutron data were processed (units conversion, detector efficiency corrections, etc.) using the Mantid software package \cite{ARNOLD2014}. Further processing, visualization and analysis, were performed using the {\sc Horace} software package \cite{EWINGS-Horace}. The linear spin wave theory analysis was done using a combination of the SpinW software package \cite{Toth_2015} and {\sc Horace}, including convolution of the instrumental resolution parameters. Fitting of the exchange and DM parameters was performed using a particle swarm optimizer \cite{pso_488968} followed by conventional least-squares optimization. Raw data are available at ref. \cite{let_data}.

\subsection{Acknowledgements}
We thank S. Nikolaev for insightful comments. This work was supported by JSPS KAKENHI Grant Number JP23KJ2165. MANA is supported by World Premier International Research Center Initiative (WPI), MEXT, Japan. The computations in this study were performed on the Numerical Materials Simulator at the NIMS. Experiments at the ISIS Neutron and Muon Source were supported by an Xpress beamtime allocation RB1690453 from the Science and Technology Facilities Council.

\bibliography{ref}

\end{document}

\title{Supplementary Information: Entangled orbital, spin, and ferroelectric orders in $p$-electron magnet CsO$_2$}

\author{Ryota Ono}
\email[Correspondence email address: ]{ryota.ono.gm@gmail.com} 
\affiliation{National Institute for Materials Science, MANA, 1-1 Namiki, Tsukuba, Ibaraki 305-0044, Japan}

\author{Ravi Kaushik}
\affiliation{Quantum Materials Theory, Italian Institute of Technology, Via Morego 30, 16163 Genova, Italy}

\author{Sergey Artyukhin}
\affiliation{Quantum Materials Theory, Italian Institute of Technology, Via Morego 30, 16163 Genova, Italy}

\author{Martin Jansen}
\affiliation{Max-Planck-Institut f\"{u}r Festk\"{o}rperforschung, D-70569 Stuttgart, Heisenbergstr. 1, Germany}

\author{Igor Solovyev}
\email[Correspondence email address: ]{SOLOVYEV.Igor@nims.go.jp}
\affiliation{National Institute for Materials Science, MANA, 1-1 Namiki, Tsukuba, Ibaraki 305-0044, Japan}

\author{Russell A. Ewings}
\email[Correspondence email address: ]{russell.ewings@stfc.ac.uk}
\affiliation{ISIS Pulsed Neutron and Muon Source, STFC Rutherford Appleton Laboratory, Harwell Campus, Didcot, Oxon OX11 0QX, United Kingdom}

\maketitle

\section{Supplementary note 1: Electronic structure}
Fig. 1 shows the electronic structure obtained from DFT calculation~\cite{QE1} and Wannier downfolding~\cite{W90} around the Fermi level. 
The density of states reveals well-isolated, nearly dispersionless band structures in CsO$_2$, indicating the highly localized nature of its molecular orbitals.
\begin{figure}[htbp]
\centering
\includegraphics[keepaspectratio, scale=0.18]{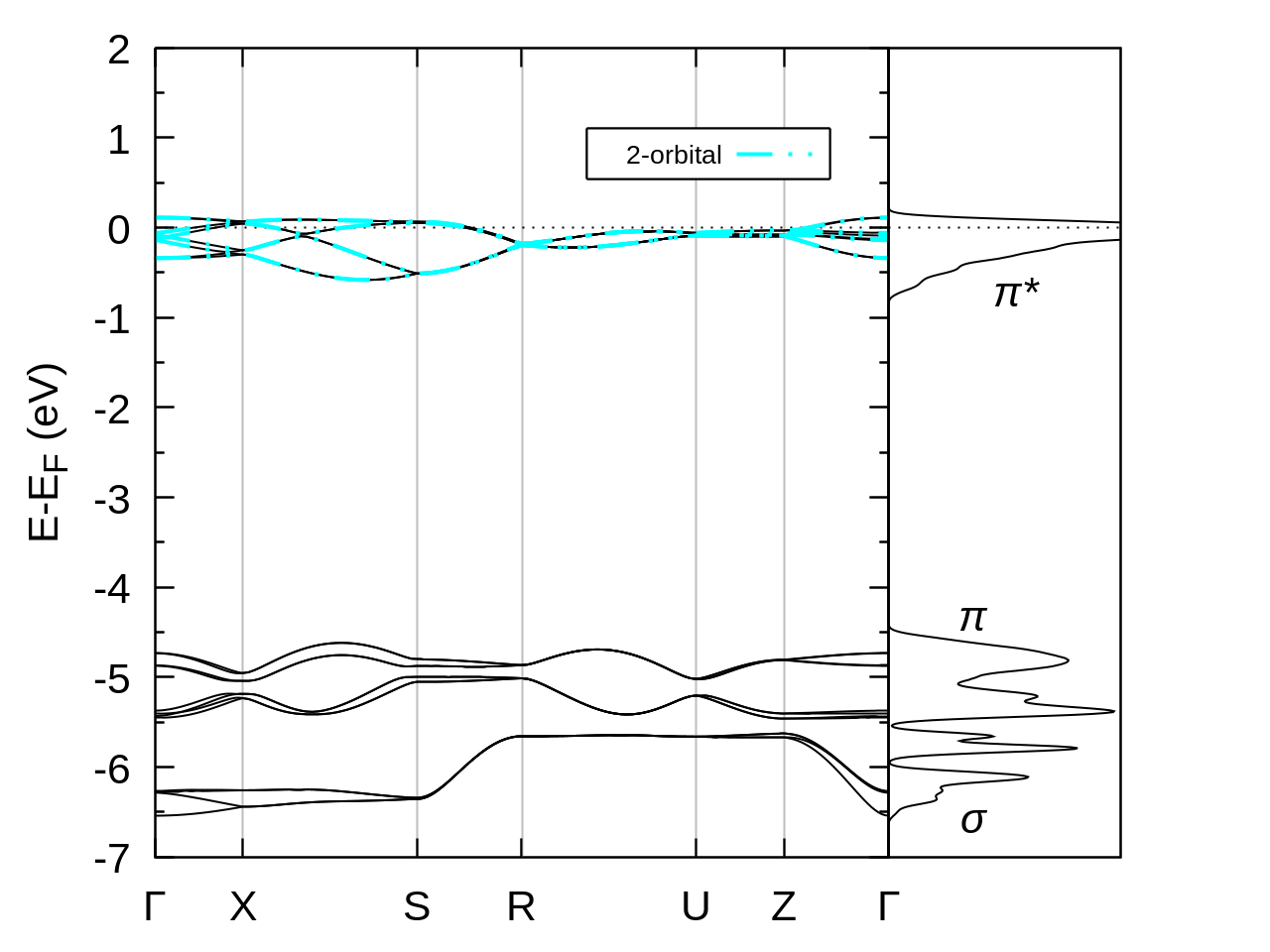}
\caption{Electronic band structure and the density of states from DFT calculation (black solid lines) and the Wannier 2-orbital model for Oxygen $\pi^*$ molecular orbital (dashed cyan lines).  }\label{fig:bands}
\end{figure}

\begin{comment}
\section{Supplementary note 2: one-electron on-site Hamiltonian}
According to the constructed Wannier functions, on-site Hamiltonian of the 2-orbital model at each site $i$ can be given as 
\begin{align*}
H_{\text{on-site}} = H_{\text{CF}}+H_{\text{SOC}}= 
    \begin{pmatrix}
     \epsilon_1 & 0  & a+b i & 0 \\
     0 & \epsilon_1 & 0 & a-b i \\
     a-b i & 0 & \epsilon_2 & 0 \\
     0 & a+b i& 0 & \epsilon_2 \\
    \end{pmatrix},
\end{align*}
where, the orbital order is taken as $\ket{\pi_1^* \uparrow}, \ket{\pi_1^* \downarrow},\ket{\pi_2^* \uparrow},\ket{\pi_2^* \downarrow} $. 
The orbital $\ket{\pi_1^*}$ and $\ket{\pi_2^*}$ are taken to be the $\pi^*$-orbital with the robes along the shorter and the longer Cs-O$_2$-Cs bond, respectively.
Actual values of the each molecule is given in Table~\ref{tab:}.
The two-folded eigenvalues are given as $ \left(\epsilon_1+\epsilon_2 \pm \sqrt{4 a^2+4 b^2+\epsilon_1^2-2 \epsilon_1 \epsilon_2 +\epsilon_2^2}\right)/2$ with the corresponding eigenstates  $\alpha_{\mp,s} \ket{\pi_1^* s}+\ket{\pi_2^* s}$, $\alpha_{\mp,s}=\left( {-\epsilon_1+\epsilon_2 \mp \sqrt{4 a^2+4 b^2+\epsilon_1^2-2 \epsilon_1 \epsilon_2+\epsilon_2^2}}\right)/ (2
(a-i s b))$, where $s=\uparrow (+1) \; \text{or} \; \downarrow (-1)$.
\end{comment}

\section{Supplementary note 2: Transition temperature}
Fig. 2 shows the specific heat in CsO$_2$ obtained from classical Monte Carlo simulations with the Metropolis algorithm applied to the full spin-Hamiltonian with various simulation sizes. 
Periodic boundary condition in all lattice vector directions are employed in each calculation.
The peak position at approximately 8\,K indicates the transition temperature, showing good agreement with the experimental value of 10\,K~\cite{Russell}. 
The calculations also reveal that the transition temperature is nearly size-independent.
\begin{figure}[htbp]
\centering
\includegraphics[keepaspectratio, scale=0.7]{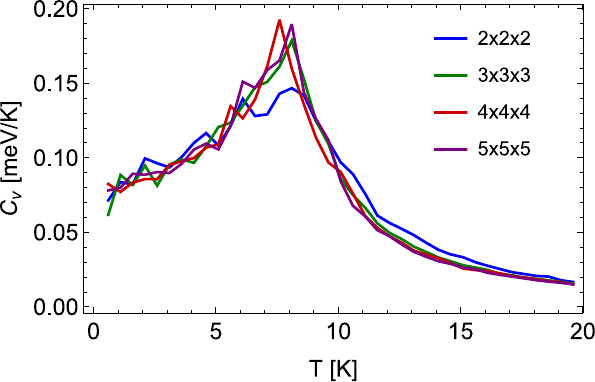}
\caption{Specific heat as calculated from classical Monte-Carlo simulations with various simulation size and periodic boundary condition.}\label{fig:Cv}
\end{figure}

\section{Supplementary note 3: Isotropic mechanism in the magnetically-induced ferroelectricity}
In this section, we discuss the symmetry property of the isotropic mechanism in the magnetically-induced ferroelectricity (the first term of Eq.~(6) in the main text).
The isotropic term is finite in non-centrosymmetric bonds $A2$, $F1$ and $F2$.
Symmetry property of these bonds are given as followings.
\begin{align}
    &\mathbf{P}_{ij}^{A2}=\left( (-1)^{i+1} P_{A2}^{x}, (-1)^{\left\lfloor (i -1)/2 \right\rfloor} P_{A2}^{y}, 0 \right), \\
    & \mathbf{P}_{ij}^{F1} = (-1)^{\left\lfloor (i -1)/2 \right\rfloor} \left( e_{ij}^{x} e_{ij}^{y} (e_{ij}^{z})^2  P_{F1}^{x} ,P_{F1}^{y},  e_{ij}^{x} e_{ij}^{y} (e_{ij}^{z})^2 P_{F1}^{z} \right) \\
     &\mathbf{P}_{ij}^{F2} = \left(  (-1)^{i\cdot j} P_{F2}^{x} , (-1)^{i\cdot j} e_{ij}^{x} \frac{i-j}{|i-j|} P_{F2}^{y},0\right), 
\end{align}
where $P_{\alpha}^{x,y,z}$ are bond dependent parameters and $\mathbf{e}_{ij} = (e_{ij}^{x},e_{ij}^{y},e_{ij}^{z}) \coloneqq sign(\hat{\epsilon_{ij}})$.
The resulting bond dependent parameters are summarized in Table~\ref{tab:param}.
In the commensurate AFM state and SF phase, this isotropic mechanism of the polarization vanishes.

\begin{table}
    \captionof{table}{Parameters of the isotropic mechanism in the magnetically-induced ferroelectricity [$\mu C /m^2$]. %calculated with 
    %$U=3.55$~eV and $J_H=0.62$~eV~\cite{Solovyev_2008}
}\label{tab:param}
\begin{ruledtabular}
\begin{tabular}{cccc}
 $\alpha$ & $P_{\alpha}^{x}$ & $P_{\alpha}^{y}$ & $P_{\alpha}^{z}$ \\
\hline
 A2  & 0.333 & 0.034 \\  
 F1  & 0.846 &  -0.028 & 0.414 \\  
 F2  & -0.013 &  1.421 \\

\end{tabular}
\end{ruledtabular}
\end{table}

\section{Supplementary note 4: Linear spin wave fitting of the inelastic neutron scattering data}

\begin{comment}
RAE to add notes here

Outline

\begin{itemize}
    \item Exposition of the challenges of fitting powder INS data with many-parameter model, especially w.r.t. local minima in cost function space, efficient optimisation routines, etc.
    \item Description of fitting procedure with just $J$.
    \item Explanation of shortcomings of this for correct magnetic structure.
    \item Plots and best fit parameters.
    \item Addition of DM terms - qualitative effect on gap.
    \item Plots and updated best fit parameters.
    \item Discussion of Gamma terms, notably inclusion of 5 more parameters to model.
\end{itemize}
\end{comment}

Powder inelastic neutron scattering data are often difficult to fit quantitatively to spin wave models \cite{BaFe2As2}, with even simple models being poorly constrained. This arises due to the spherical averaging intrinsic to powder measurements, which acts to wash out fine structure. The method is often also insensitive to differences in spin wave dispersions between the Brillouin zone center and high symmetry points with the same or similar magnitude of wavevector. For complex models with many parameters, such as is required for the case of CsO$_{2}$, this leads to an extremely challenging situation, with many possible permutations of model parameters giving rise to near identical fit quality, characterised by the goodness-of-fit parameter $\chi^{2}$. Put another way, the $\chi^{2}$ landscape has high dimensionality and contains many local minima in which conventional steepest descent least squares parameter optimisation tools can get stuck. Even with non-gradient based methods such as particle swarm \cite{pso_488968}, if many parameter combinations yield near identical values of $\chi^{2}$ there is no guarantee that the true best fit will be found.

To explore the parameter space for CsO$_{2}$ as best as possible, we used the particle swarm method as implemented in the Matlab global optimisation toolbox \cite{Op}. A swarm size of 2000 was chosen, which is substantially larger than the default option of ten times the number of parameters, in order to explore the parameter space as thoroughly as possible while avoiding excessive computational cost. A particular challenge particular to this system is that large parts of parameter space are `forbidden', in the sense that they are inconsistent with the known magnetic structure. On cannot deal with this via a simple approach of setting $\chi^{2}$ to be very large in these regions, since this acts as a vertical `wall' between regions of stability in parameter space. Instead, in these regions $\chi^{2}$ was assigned a value proportional to the size of the imaginary eigenvalues in the spin wave calculation. Even so, this does not permit particles in the swarm to move freely around parameter space. So, we are unable to be confident that the best fits found via the particle swarm method are true global optima. However, our experience of running the swarm optimisation iteratively with different starting parameters was that qualitatively similar best fits were found each time.

The minimal model that can be fitted to the spin wave data includes the five principal exchange parameters, and the Dzyaloshinskii-Moriya terms. The latter are an absolute requirement to stabilize the known magnetic structure, which exhibits a small antiferromagnetic canting of the moments towards the $a$-axis. Including the $\Gamma$ terms generally improves the best fits slightly, although many different combinations of parameters yielded very similar fit quality. Examples of typical good fits are shown in the figures below, with the fit parameters shown in table \ref{tab:fitpars}. The first two rows correspond to fits when the $\Gamma$ parameters were fixed to zero, whereas the last two rows show the results when they were allowed to be non-zero.

%Mapping
% JA1 = J3
% JA2 = J2
% JA3 = J5
% JF1 = J4
% JF2 = J1

\begin{table}
    %\captionof{table}
    \caption{Goodness-of-fit parameter, and corresponding parameters, for various different attempts at parameter optimisation. All model parameters are given in units of meV. 
    %Note that only the $\Gamma_\alpha^{||}$ parameters are given explicitly, because $\Gamma_\alpha^{zz} = -2\Gamma_\alpha^{||}$.
    }\label{tab:fitpars}
\begin{ruledtabular}
\begin{tabular}{cccccccccccccc}
 $\chi^{2}$ & $J_{\rm{A1}}$ & $J_{\rm{A2}}$ &  $J_{\rm{A3}}$ & $J_{\rm{F1}}$ & $J_{\rm{F2}}$ & $d_{\rm{A2}}$ & $d_{\rm{F1}}$ & $d_{\rm{F2}}$ & $\Gamma_{\rm{A1}}^{||}$ & $\Gamma_{\rm{A2}}^{||}$ & $\Gamma_{\rm{A3}}^{||}$ & $\Gamma_{\rm{F1}}^{||}$ & $\Gamma_{\rm{F2}}^{||}$ \\
\hline
  1.02 & 5.73 & 0.73 & 0.44 & -1.11 & -1.01 & 0.0 & 1.26 & 2.82 & 0 & 0 & 0 & 0 & 0  \\
  0.95 & 5.36 & 0.55 & 0.19 & -1.17 & -0.97 & 0.0 & 1.48 & 3.32 & 0 & 0 & 0 & 0 & 0  \\  
  0.91 & 5.40 & 0.45 & 0.10 & -1.13 & -0.94 & 0.0 & 1.03 & 2.43 & 0.0 & 0.002 & 0.0 & -0.001 & 0.004  \\
  0.93 & 5.0 & 0.41 & 0.04 & -1.31 & -1.27 & 0.0 & 1.29 & 3.75 & 0.01 & 0.09 & 0.11 & -0.22 & -0.52  \\  
\end{tabular}
\end{ruledtabular}
\end{table}

\begin{figure}[!h]
\centering
    \includegraphics[scale=0.6]{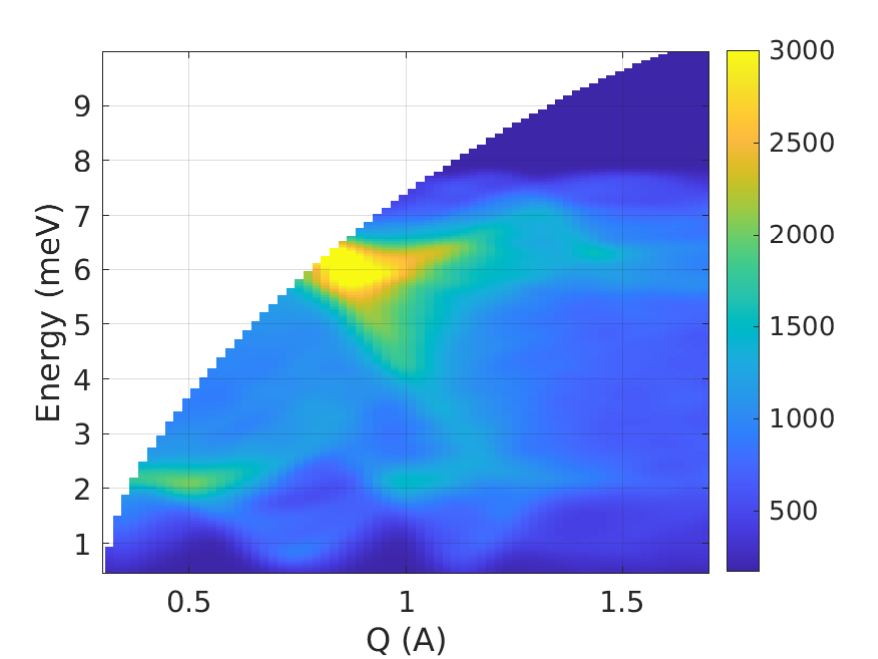}
    \caption{Fit to experimental data, with the parameters given in the first row of table \ref{tab:fitpars}.}
    \label{f:Ei11_62}
\end{figure}

\begin{figure}[!h]
\centering
    \includegraphics[scale=0.6]{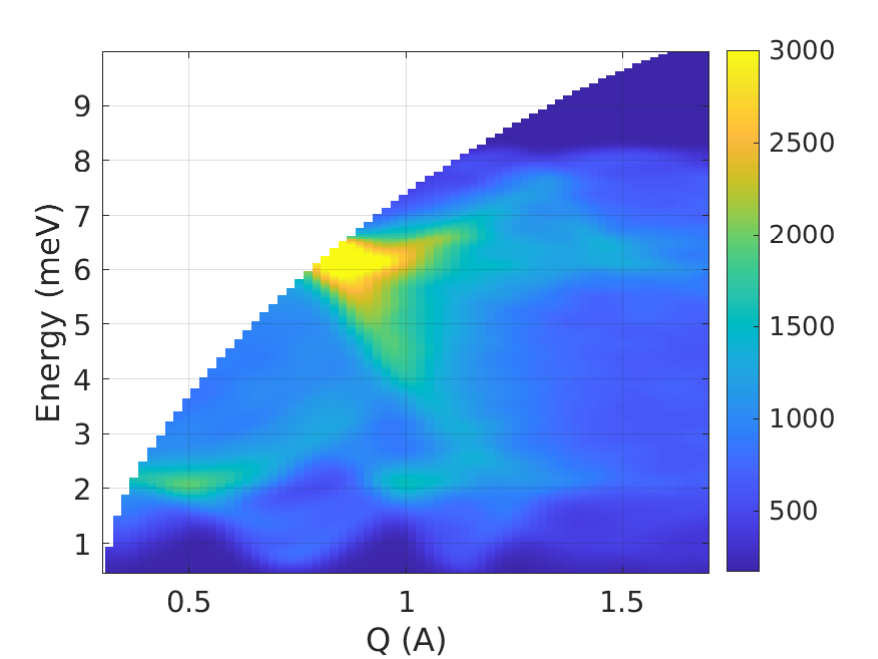}
    \caption{Fit to experimental data, with the parameters given in the second row of table \ref{tab:fitpars}.}
    \label{f:Ei11_6_5}
\end{figure}

\begin{figure}[!h]
\centering
    \includegraphics[scale=0.6]{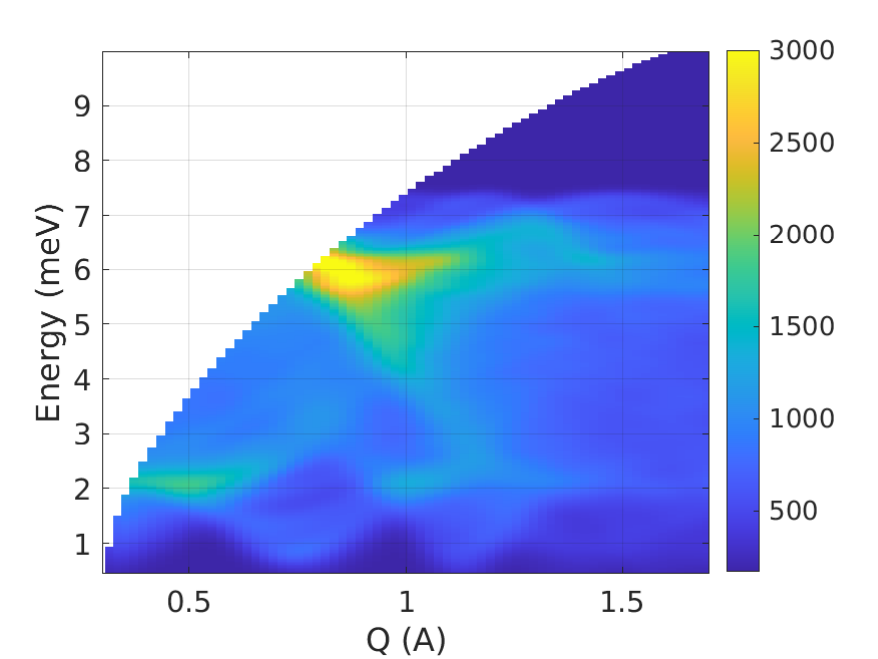}
    \caption{Fit to experimental data, with the parameters given in the third row of table \ref{tab:fitpars}.}
    \label{f:Ei11_7b3}
\end{figure}

\begin{figure}[!h]
\centering
    \includegraphics[scale=0.6]{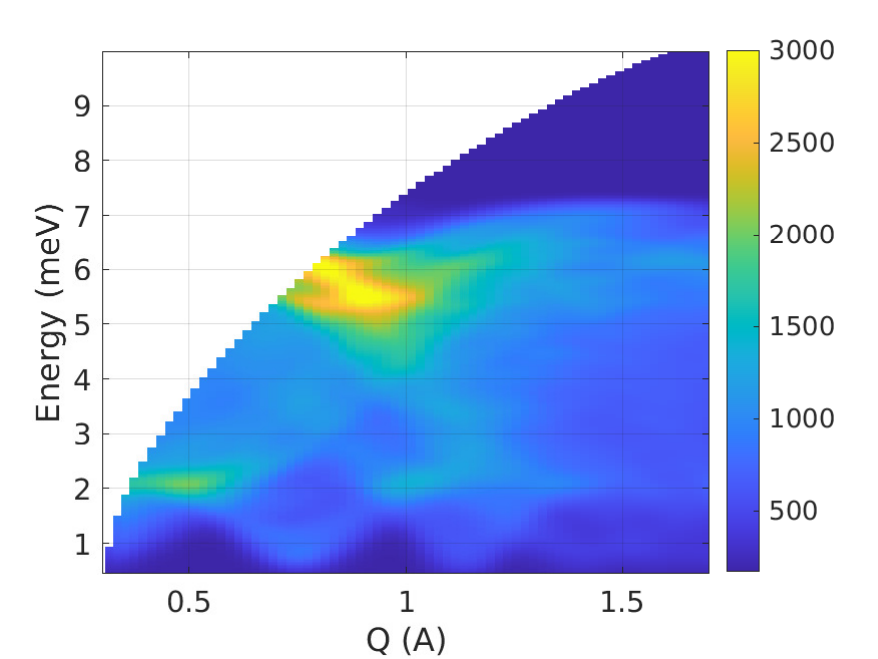}
    \caption{Fit to experimental data, with the parameters given in the fourth row of table \ref{tab:fitpars}.}
    \label{f:Ei11_84}
\end{figure}

\bibliography{ref_sup.bib}

%% file: preamble.tex
\usepackage{amsthm}
\usepackage{mathtools}
\usepackage{physics}
\usepackage{xcolor}
\usepackage{graphicx}
\usepackage[left=23mm,right=13mm,top=35mm,columnsep=15pt]{geometry} 
\usepackage{adjustbox}
\usepackage{placeins}
\usepackage[T1]{fontenc}
\usepackage{lipsum}
\usepackage{csquotes,comment}